\documentclass[useAMS,usenatbib]{mn2e}
\usepackage{graphicx}
\usepackage{color}
\usepackage{amssymb}
\usepackage{amsmath}
\usepackage{upgreek}
\usepackage{subfigure}
\usepackage{rotating}
\usepackage{pdflscape}
\usepackage{bm} 
\usepackage{mathtools}

\usepackage{comment}
\graphicspath{{images/}}

\title[{Stellar cosmic rays at the young Earth}]{{Stellar versus Galactic: The intensity of cosmic rays at the evolving Earth and young exoplanets around Sun-like stars}}
\author[Rodgers-Lee et al.]{D. Rodgers-Lee$^{1}$\thanks{E-mail:
drodgers@tcd.ie}, A. M. Taylor$^{2}$, A. A. Vidotto$^{1}$ and  T. P. Downes$^{3}$\\ 
$^{1}$ School of Physics, Trinity College Dublin, University of Dublin, College Green, Dublin 2, D02 PN40, Ireland \\
$^{2}$ DESY, D-15738 Zeuthen, Germany \\
$^{3}$ Centre for Astrophysics \& Relativity, School of Mathematical Sciences, Dublin City University, Glasnevin, D09 W6Y4, Ireland
}

\begin{document}
\date{ Accepted 2021 March 23. Received 2021 March 9; in original form 2021 January 22.}
\pagerange{\pageref{firstpage}--\pageref{lastpage}} \pubyear{xxxx}
\maketitle

\label{firstpage}

\begin{abstract} 
Energetic particles, such as stellar cosmic rays, produced at a heightened rate by active stars (like the young Sun) may have been important for the origin of life on Earth and other exoplanets. Here we compare, as a function of stellar rotation rate ($\Omega$), contributions from two distinct populations of energetic particles: stellar cosmic rays accelerated by impulsive flare events and Galactic cosmic rays. We use a 1.5D stellar wind model combined with a spatially 1D cosmic ray transport model. We formulate the evolution of the stellar cosmic ray spectrum as a function of stellar rotation. The maximum stellar cosmic ray energy increases with increasing rotation i.e., towards more active/younger stars. We find that stellar cosmic rays dominate over Galactic cosmic rays in the habitable zone at the pion threshold energy for all stellar ages considered ($t_*=0.6-2.9\,$Gyr). However, even at the youngest age, $t_*=0.6\,$Gyr, we estimate that $\gtrsim\,80$MeV stellar cosmic ray fluxes may still be transient in time. At $\sim$1\,Gyr when life is thought to have emerged on Earth, we demonstrate that stellar cosmic rays dominate over Galactic cosmic rays up to $\sim$4\,GeV energies during flare events. Our results for $t_*=$0.6\,Gyr ($\Omega = 4\Omega_\odot$) indicate that $\lesssim$GeV stellar cosmic rays are advected from the star to 1\,au and are impacted by adiabatic losses in this region. The properties of the inner solar wind, currently being investigated by the Parker Solar Probe and Solar Orbiter, are thus important for accurate calculations of stellar cosmic rays around young Sun-like stars.
\end{abstract}


\section{Introduction}
\label{sec:intro}

There is much interest in determining the conditions, such as the sources of ionisation for exoplanetary atmospheres, that were present in the early solar system when life is thought to have begun on Earth \citep[at a stellar age of $\sim$1\,Gyr,][]{mojzsis_1996}. This allows us to postulate what the important factors that led to life here on Earth were. These studies can then be extended to young exoplanets around solar-type stars whose atmospheres may be characterised in the near future by upcoming missions, such as the James Webb Space Telescope \citep[JWST,][]{gardner_2006,barstow_2016}.

Cosmic rays represent a source of ionisation \citep{rimmer_2013} and heating \citep{roble_1987, glassgold_2012} for exoplanetary atmospheres. In this paper we compare the contributions from two distinct populations of energetic particles: stellar cosmic rays accelerated by their host stars and Galactic cosmic rays. Galactic cosmic rays reach Earth after travelling through the heliosphere \citep[see review by][]{potgieter_2013}. These cosmic rays originate from our own Galaxy and constitute a reservoir of relativistic particles in the interstellar medium (ISM) that diffuse through the magnetised solar wind in a momentum-dependent way. 

The propagation of Galactic cosmic rays through the stellar winds of younger solar-type stars has previously been studied \citep{svensmark_2006,cohen_2012,rodgers-lee_2020b}. The intensity of Galactic cosmic rays that reached the young Earth ($\sim$1\,Gyr old) is thought to be much reduced in comparison to present-day observed values. This is due to the increased velocity and magnetic field strength present in the stellar wind of a young ($\sim$Gyr old) solar-type star in comparison to the present-day solar wind (assuming that the turbulence properties of the wind remain constant with stellar age). Similar to what has been estimated to occur for Galactic cosmic rays, the changing physical conditions of the stellar wind throughout the life of a Sun-like star will affect the propagation of stellar cosmic rays. The propagation of Galactic cosmic rays through the astrospheres\footnote{The more general term for the heliosphere of other stellar systems} of a number of M dwarf stars has also recently been considered \citep{herbst_2020,mesquita_2021}.

Here, we investigate the intensity of solar, or more generally stellar, cosmic rays as they propagate through the wind of a solar-type star throughout the star's life, particularly focusing on the intensity at the orbital distance of Earth. Solar/stellar cosmic rays are also known as solar/stellar energetic particles. We focus on solar-type stars so that our results can also be interpreted in the context of the young Sun and the origin of life on Earth.

Cosmic rays are thought to be important for prebiotic chemistry and therefore may play a role in the origin of life \citep{dartnell_2011,rimmer_2014,airapetian_2016, dong_2019}. Cosmic rays may also result in observable chemical effects in exoplanetary atmospheres by leading to the production of molecules such as $\text{NH}_4^+,\, \mathrm{H_3^+}$ and $\mathrm{H_3O^+}$ \citep{helling_2019,barth_2020}. In addition to this, cosmic rays may lead to the production of \emph{fake} biosignatures via chemical reactions involving ${\rm NO_x}$ \citep{grenfell_2013}. Biosignatures are chemical signatures that are believed to be the chemical signatures of life, such as molecular oxygen \citep{meadows_2018}. Thus, in order to interpret upcoming observations which will focus on detecting biosignatures we must constrain the contribution of cosmic rays to fake biosignatures.

An interesting aspect that we focus on in this paper is the fact that the intensity and momentum of stellar cosmic rays accelerated by a solar-type star most likely increase for younger stars due to their stronger stellar magnetic fields, unlike the Galactic cosmic ray spectrum which is assumed to remain constant with time. The present day Sun, despite being an inactive star, has been inferred to accelerate particles to GeV energies in strong solar flares \citep{ackermann_2014,ajello_2014,kafexhiu_2018}. There is also evidence from cosmogenic nuclides to suggest that large solar energetic particle events occurred even in the last few thousand years \citep{miyake_2019}. Thus, it is very likely that a younger Sun would accelerate particles at a higher rate, and to higher energies, due to the stronger magnetic field strengths observed for young Sun-like stars \citep[e.g.][]{johns-krull_2007,hussain_2009,donati_2014} and at a more continuous rate due to an increased frequency of stellar flares \citep{maehara_2012,maehara_2015}. A scaled up version of a large solar energetic particle event is often assumed as representative of stellar cosmic rays around main sequence M dwarf stars or young pre-main sequence solar-type stars. The work presented in this paper builds upon this research and aims to contribute towards a clearer and broader understanding of the spectral shape and intensity of stellar cosmic rays which reach exoplanets around solar-type stars as a function of age. 

There has also been a significant amount of research concerning the propagation of stellar cosmic rays through the magnetospheres and atmospheres of close-in exoplanets around M dwarf stars \citep{segura_2010,tabataba-vakili_2016}, as well as comparisons with Galactic cosmic rays \citep{griessmeier_2015}. \citet{atri_2020} also investigated the surface radiation dose for exoplanets resulting from stellar cosmic ray events starting at the top of an exoplanetary atmosphere considering an atmosphere with the same composition as Earth's. Our results can be used in the future as an input for these types of studies.

In this paper we compare the relative intensities of stellar and Galactic cosmic rays of different energies as a function of stellar age. This allows us to estimate the age of a solar-mass star when the intensities of stellar and Galactic cosmic rays are comparable, at a given energy. This also depends on the orbital distance being considered. Here we focus mainly on the habitable zone of a solar-mass star where the presence of liquid water may be conducive to the development of life \citep{kasting_1993}.

Previous studies have estimated the intensity of Galactic cosmic rays at $\sim$1\,Gyr when life is thought to have begun on Earth. Stellar cosmic rays have separately been considered in the context of T-Tauri systems \citep{rab_2017,rodgers-lee_2017,rodgers-lee_2020, fraschetti_2018,offner_2019} and more generally in star-forming regions \citep[see][for a recent review]{padovani_2020}. \citet{fraschetti_2019} also investigated the impact of stellar cosmic rays for the Trappist-1 system. \citet{scheucher_2020} focused on the chemical effect of a large stellar energetic particle event on the habitability of Proxima Cen b. However, the propagation of stellar cosmic rays through stellar systems has not yet been investigated as a function of stellar rotation rate or at the potentially critical time when the Sun was $\sim$1\,Gyr old. We also compare the relative intensities of stellar and Galactic cosmic rays for the HR2562 system \citep{konopacky_2016} which we focused on previously in \citet{rodgers-lee_2020b}.

The paper is structured as follows: in Section\,\ref{sec:form} we describe the details of our model and the properties that we have adopted for the stellar cosmic rays. In Section\,\ref{sec:results} we present and discuss our results in relation to the young Sun and the young exoplanet, HR 2562b. Finally, we present our conclusions in Section\,\ref{sec:conclusions}.

\section{Formulation}
\label{sec:form}

In this section we motivate our stellar cosmic ray spectrum as a function of stellar rotation rate. We also briefly describe the cosmic ray transport model and stellar wind model that we use \citep[previously presented in][]{rodgers-lee_2020b}.

\subsection{From solar to stellar cosmic rays}
The present-day Sun is the only star for which we can directly detect solar cosmic rays and determine the energy spectrum of energetic particles arriving to Earth. We use these observations of the present-day Sun to guide our estimate for the stellar cosmic ray spectrum of solar-type stars of different ages, which are representative of the Sun in the past. The shape of the energy spectrum and the overall power in stellar cosmic rays are the two quantities required to describe a stellar energetic particle spectrum (Section\,\ref{subsec:qinj}).

Solar energetic particle (SEP) events can broadly be divided into two categories known as gradual and impulsive events \citep[][for instance]{reames_2013,klein_2017}. Gradual SEP events are thought to be mainly driven by the acceleration of particles at the shock fronts of coronal mass ejections (CMEs) as they propagate. These events produce the largest fluences of protons at Earth. Impulsive SEP events are associated with flares close to the corona of the Sun and while they result in lower proton fluences at Earth they occur more frequently than gradual events. The terms `gradual' and `impulsive' refer to the associated X-ray signatures. 

While gradual SEP events associated with CMEs produce the largest fluences of protons detected at Earth it is unclear what energies the CMEs would have and how frequently they occur for younger solar-type stars \citep{aarnio_2012, drake_2013,osten_2015}. Very large intensities of stellar cosmic rays associated with very energetic, but infrequent, CMEs may simply wipe out any existing life \citep{cullings_2006,atri_2020} on young exoplanets rather than helping to kick start it. On the other hand lower intensities of stellar cosmic rays, associated with impulsive flare events, at a more constant rate may be more of a catalyst for life \citep{atri_2016,lingam_2018,dong_2019}. In the context of the potential impact of stellar energetic particles on exoplanetary atmospheres we restrict our focus here to protons. This is because only protons can be accelerated to $\sim$GeV energies, rather than electrons which suffer from energy losses.

Many white light (referring to broad-band continuum enhancement, rather than chromospheric line emission, for instance) flares have been detected by the Kepler mission \citep{koch_2010}. An increase in the frequency of super-flares (bolometric flare energies of $>10^{33}\,$erg) with increasing stellar rotation (i.e. younger stars) has also been found \citep{maehara_2015}. Some of the most energetic white light flares are from pre-main sequence stars in the Orion complex detected in the Next Generation Transit Survey \citep{jackman_2020}. Since SEP events often have associated optical and X-ray emission, the detection of very energetic white light flares from younger stars/faster rotators is presumed to lead to a corresponding increase in X-rays. Indeed, young stars are known to be stronger X-ray sources in comparison to the Sun \citep{feigelson_2002}. Therefore, it seems likely that stars younger than the Sun will also produce more stellar energetic particles than the present day Sun \citep[see][for a recent estimate of stellar proton fluxes derived using the empirical relation between stellar effective temperature and starspot temperature]{herbst_2020b}. 

Somewhat surprisingly, given the number of superflares detected with Kepler, there have only been a small number of stellar CME candidate events \citep{argiroffi_2019,moschou_2019,vida_2019,leitzinger_2020}. To investigate the possibility that stellar CMEs are not as frequent as would be expected by extrapolating the solar flare-CME relation \citep{aarnio_2012,drake_2013,osten_2015}, \citet{alvarado-gomez_2018} illustrated using magnetohydrodynamic simulations that a strong large-scale stellar dipolar magnetic field (associated with fast rotators) may suppress CMEs below a certain energy threshold. Another line of argument discussed in \citet{drake_2013} suggests that the solar flare-CME relationship may not hold for more active stars because the high CME rate expected for active stars (obtained by extrapolating the solar flare-CME relationship) would lead to very high stellar mass-loss rates. This has not been found for mass-loss rates inferred from astrospheric Ly$\alpha$ observations \citep{wood_2002,wood_2004,wood_2014} or from transmission spectroscopy, coupled to planetary atmospheric evaporation and stellar wind models \citep{vidotto_2017b}. At the same time the number of stars with estimates for their mass-loss rates remains small. 

Thus, as a first estimate for the intensity of stellar energetic particles impinging on exoplanetary atmospheres we consider flare-accelerated protons that we inject close to the surface of the star. We do not consider stellar energetic particles accelerated by shocks associated with propagating CMEs due to the current lack of observational constraints for the occurence rate and energy of stellar CMEs as a function of stellar age. Our treatment of the impulsive stellar cosmic ray events is described in the following section.

Our investigation treats the injection of stellar cosmic rays as continuous in time during a given epoch of a star's life. Two key factors here that control the applicability of such an assumption are that young solar-type stars (i.e. fast rotators) are known to flare more frequently than the present-day Sun, and their associated flare intensity at a given frequency is more powerful \citep{salter_2008,maehara_2012,maehara_2015}. In order to focus on stellar cosmic rays injected at such a heightened rate and power, and thus can be treated as continuous in time, we restrict our results to stellar rotation rates greater than the rotation rate of the present-day Sun. We discuss this assumption in more detail in Section\,\ref{subsec:injection_assumption}.

\subsection{Transport equation for stellar cosmic rays}
To model the propagation of stellar cosmic rays from a solar-type star out through the stellar system we solve the 1D cosmic ray transport equation \citep[derived by][for the modulation of Galactic cosmic rays in the solar system]{parker_1965}, assuming spherical symmetry. We use the same numerical code as presented in \citet{rodgers-lee_2020b} which includes spatial diffusion, spatial advection and energy losses due to momentum advection of the cosmic rays. The 1D transport equation is given by
\begin{equation}
\frac{\partial f}{\partial t} = \nabla\cdot(\kappa\nabla f)-v\cdot\nabla f +\frac{1}{3}(\nabla\cdot v)\frac{\partial f}{\partial \mathrm{ln}p} + Q,
\label{eq:f}
\end{equation}
\noindent where $f(r,p,t)$ is the cosmic ray phase space density, $\kappa(r,p,\Omega)$ is the spatial diffusion coefficient, $v(r,\Omega)$ is the radial velocity of the stellar wind and $p$ is the momentum of the cosmic rays (taken to be protons)\footnote{There was a typographical error in Eq.\,1 of \citet{rodgers-lee_2020b} where the $v\cdot\nabla f$ term was expressed as $\nabla \cdot(vf)$.}. $Q$ is defined as
\begin{equation}
Q(r_\mathrm{inj},p,\Omega)=\frac{1}{4\pi p^2}\frac{d\dot N}{d^3x\,dp}, \label{eq:q}
\end{equation}
which represents the volumetric injection of stellar cosmic rays per unit time and per unit interval in momentum, injected at a radius of $r_\mathrm{inj} \sim1.3R_\odot$. $\dot N=dN/dt$ is the number of particles injected per unit time. $Q$ varies as a function of stellar rotation rate, $\Omega$. The details of how we treat the injection rate are discussed in Section\,\ref{subsec:qinj}.  The numerical scheme and the simulation set-up are otherwise the same as that of \citet{rodgers-lee_2020b}.
 
We assume an isotropic diffusion coefficient, which varies spatially by scaling with the magnetic field strength of the stellar wind (and on the level of turbulence present in it) and depends on the cosmic ray momentum \citep[see Eq.\,3 of][]{rodgers-lee_2020b}. Here, for simplicity we take the level of turbulence to be independent of stellar rotation rate using the same value as motivated in \citet{rodgers-lee_2020b}. Spatial advection and the adiabatic losses of the cosmic rays depend on the velocity and divergence of the stellar wind. 

In the context of the modulation of {\it Galactic} cosmic rays spatial and momentum advection collectively result in the suppression of the local interstellar spectrum (LIS) of Galactic cosmic rays that we observe at Earth. For {\it stellar} cosmic rays the situation is slightly different due to their place of origin in the system. Stellar cosmic rays still suffer adiabatic losses as they travel through the stellar wind via the momentum advection term, but the spatial advective term now merely advects the stellar cosmic rays out through the solar system. Spatial advection only operates as a loss term if the stellar cosmic rays are advected the whole way through and out of the stellar system.

\subsection{Stellar wind model}
In our model the stellar wind of a Sun-like star is launched due to thermal pressure gradients and magneto-centrifugal forces in the hot corona overcoming stellar gravity \citep{weber_1967}. The wind is heated as it expands following a polytropic equation of state. Stellar rotation is accounted for in our model leading to (a) angular momentum loss via the magnetic field frozen into the wind and (b) the development of an azimuthal component of an initially radial magnetic field. We use the same stellar wind model as in \citet{rodgers-lee_2020b} that is presented in more detail in \citet{carolan_2019}. Our 1.5D polytropic magneto-rotator stellar wind model \citep{weber_1967} is modelled with the Versatile Advection Code \citep[VAC,][]{toth_1996,johnstone_2015a} and here we provide a brief summary of it. 

By providing the stellar rotation rate, magnetic field, density and temperature at the base of the wind as input parameters for the model we are able to determine the magnetic field strength, velocity and density of the stellar wind as a function of orbital distance out to 1\,au. Beyond 1\,au the properties of the stellar wind are extrapolated out to the edge of the astrosphere as discussed in Section\,2.3.2 of \citet{rodgers-lee_2020b}.
The main inputs for the stellar wind model that we varied in \citet{rodgers-lee_2020b} to retrieve the wind properties of a solar-type star for different ages were (a) the stellar rotation rate itself \citep[derived from observations of solar-type stars of different ages,][]{gallet_2013}, (b) the stellar magnetic field strength \citep[using the scaling law presented in][]{vidotto_2014}, (c) the base temperature \citep[using the empirical relationship with stellar rotation rate from][]{ofionnagain_2018} and (d) the base density of the stellar wind \citep[following][]{ivanova_2003}. In the stellar wind model the base, or launching point, of the wind is chosen to be $1\,R_\odot$ for the instances in time that we investigate.
Table\,\ref{table:sim_parameters} provides the stellar rotation rates/ages that we consider here. The corresponding stellar surface magnetic field strength, base density and temperature that we use are given in Table 1 of \citet{rodgers-lee_2020b}. Generally, the magnetic field strengths and velocities of the stellar wind increase with increasing stellar rotation rate.
\subsection{Stellar cosmic ray spectrum}
\label{subsec:qinj}

We assume a continuous injection spectrum for the stellar cosmic rays such that $d\dot N/dp \propto dN/dp \propto p^{-\alpha}$. We adopt a power law index of $\alpha = 2$ which in the limit of a strong non-relativistic shock is representative of diffusive shock acceleration \citep[DSA, as first analytically derived by][]{krymskii_1977,bell_1978,blandford_1978} or compatible with acceleration due to magnetic reconnection. We relate the total injected kinetic power in stellar cosmic rays, $L_\text{CR}$ (which we define in Section\,\ref{subsubsec:pcr}), to $d\dot N/dp$ in the following way
\begin{eqnarray}
L_\mathrm{CR} &=& \int \limits_0^\infty \frac{d\dot N}{dp} T(p)dp \label{eq:ratea} \\
 &\approx& \left.\frac{d\dot N}{dp}\right |_{2mc} \int \limits_{p_\text{0}}^{p_\mathrm{M}}  \left(\frac{p}{2mc} \right)^{-\alpha}  e^{(-p/p_\text{max})}T(p)dp, 
\label{eq:rateb}
\end{eqnarray} 

\noindent where $m$ is the proton mass, $c$ is the speed of light and $T(p) = mc^2( \sqrt{1+(p/mc)^2} - 1 )$ is the kinetic energy of the cosmic rays. $p_\mathrm{max}$ is the maximum momentum that the cosmic rays are accelerated to which is discussed further in Section\,\ref{subsubsec:pb}. The logarithmically spaced momentum bins for the cosmic rays are given by $p_k = \mathrm{exp}\{k\times\mathrm{ln}(p_M/p_0)/(M-1) + \mathrm{ln}\,p_0\}$  for $k=0,...,M$ with $M=60$. The extent of the momentum grid that we consider ranges from $p_0=0.15\,\mathrm{GeV}/c $ to $p_{\rm M}=100\,\mathrm{GeV}/c$, respectively.

We have normalised the power law in Eq.\,\ref{eq:rateb} to a momentum of $2mc$ since this demarks the part of integrand which dominates the integral (for spectra in the range $2 <\alpha < 3$ of primary interest to us). To illustrate this, following \citet{drury_1989}, Eq.\,\ref{eq:ratea} can be approximated as
\begin{flalign}
\int \limits_0^\infty \frac{d\dot N}{dp} T(p)dp &\approx& \left.\frac{d\dot N}{dp}\right |_{2mc} \int \limits_{p_0}^{2mc}  \left(\frac{p}{2mc} \right)^{-\alpha}  e^{(-p/p_\text{max})}\frac{p^{2}}{2m} dp\nonumber\\
&+&   \left.\frac{d\dot N}{dp}\right |_{2mc} \int \limits_{2mc}^{p_\text{M}} \left(\frac{p}{2mc} \right)^{-\alpha}  e^{(-p/p_\text{max})}pc\,dp,
\label{eq:rate2}
\end{flalign}
which has split the integral into a non-relativistic and relativistic component given by the first and second term on the righthand side of Eq.\,\ref{eq:rate2}, respectively. Eq.\,\ref{eq:rate2} implicitly assumes that $p_0\ll2mc$. For $\alpha=2$, Eq.\,\ref{eq:rate2} can be estimated as
\begin{eqnarray}
L_{\rm CR}\approx \left.p^2\frac{d\dot N}{dp}\right |_{2mc}\left[1-\frac{p_\mathrm{0}}{2mc}+\mathrm{ln}\left(\frac{p_\mathrm{max}}{2mc}\right)\right]
\label{eq:pinj}
\end{eqnarray}
Thus, considering $p_\mathrm{max}\sim 3\,\mathrm{GeV}/c$ indicates that the first and last term contribute approximately equally in Eq.\,\ref{eq:pinj}. $p^2\dfrac{d\dot N}{dp}|_{2mc}$ is chosen to normalise the integral to the required value of $L_\mathrm{CR}$.  

\subsubsection{Spectral break as a function of stellar rotation rate}
\label{subsubsec:pb}
The maximum momentum of the accelerated cosmic rays, $p_\mathrm{max}$, is another important parameter that we must estimate as a function of $\Omega$. It physically represents the maximum momentum of stellar cosmic rays that we assume the star is able to efficiently accelerate particles to. Both DSA and magnetic reconnection rely on converting magnetic energy to kinetic energy. The magnetic field strength of Sun-like stars is generally accepted to increase with increasing stellar rotation rate, or decreasing stellar age \citep{vidotto_2014,folsom_2018}. This indicates that more magnetic energy would have been available at earlier times in the Sun's life or for other stars rotating faster/younger than the present-day Sun to produce stellar cosmic rays. 

Therefore we evolve $p_\mathrm{max}$ as a function of stellar magnetic field strength. In our model this effectively means that the maximum injected momentum evolves as a function of stellar rotation rate. We assume that 
\begin{equation}
p_\mathrm{max}(\Omega) = p_\mathrm{max,\odot}\left(\frac{B_*(\Omega)}{B_\mathrm{*,\odot}}\right).
\label{eq:pb}
\end{equation}
We chose $p_\mathrm{max,\odot} = 0.2\,\mathrm{GeV}/c$, corresponding to a kinetic energy of $T_\mathrm{max}=20\,$MeV \citep[][for instance, report impulsive stellar energetic particle events with kinetic energies $\gtrsim50$\,MeV]{kouloumvakos_2015}. We also investigate the effect of assuming $p_\mathrm{max,\odot} = 0.6\,\mathrm{GeV}/c$ (corresponding to $T_\mathrm{max}=200\,$MeV). The values for $p_\mathrm{max}(\Omega)$ are given in Table\,\ref{table:sim_parameters}. The maximum value that we use is 3.3GeV$/c$ for a stellar rotation rate of $3.5\Omega_\odot$ at $t_*=600$\,Myr using $p_\mathrm{max,\odot} = 0.6\,\mathrm{GeV}/c$. In comparison, \citet{padovani_2015} estimated a maximum energy of $\sim$30\,GeV for the acceleration of protons at protostellar surface shocks for $t_*\lesssim 1\,$Myr. The maximum momentum of 3.3\,GeV$/c$ that we adopt corresponds to a surface average large-scale magnetic field of $\sim 8$\,G at 600\,Myr. If we investigated younger stellar ages when it would be reasonable to adopt an average large-scale stellar magnetic field of $\sim 80$\,G then we would also find a maximum energy of $\sim$30\,GeV. 

Eq.\,\ref{eq:pb} is motivated by the Hillas criterion \citep{hillas_1984} which estimates that the maximum momentum achieved at a shock can be obtained using 
\begin{flalign}
p_\text{max}c\sim q\beta_sB_{\rm s}R_{\rm s}=\frac{1}{c}\left(\frac{v}{100{\rm \,km\,s^{-1}}}\right)\left(\frac{B}{\rm 10\,G}\right)\left(\frac{R}{10^{9}\,{\rm cm}}\right){\rm GeV}\nonumber\\ 
\label{eq:hillas}
\end{flalign} 
\noindent where $\beta_s=v_s/c$ and $v_s$ is the characteristic velocity associated to the scattering agent giving rise to acceleration (eg. shock velocity or turbulence velocity), $B_{\rm s}$ is the magnetic field strength within the source, and $R_{\rm s}$ is the size of the source region. If we assume that the size of the emitting region (a certain fraction of the Sun's surface) and the characteristic velocity do not change as a function of stellar rotation rate we simply obtain $p_\text{max}\propto B_{\rm s}$ as adopted in Eq.\,\ref{eq:pb}.

Indeed, high energy $\gamma-$ray observations from {\it Fermi}-LAT found that for strong solar flares the inferred proton spectrum, located close to the surface of the Sun, displays a high maximum kinetic energy break of $\gtrsim 5\,$GeV \citep{ackermann_2014,ajello_2014}. Generally, the spectral break occurs at lower kinetic energies, or momenta, for less energetic but more frequent solar flares. Since the power law break in the SEP spectrum shifts to higher energies during strong solar flares this is a good indicator that $p_\mathrm{max}$ is likely to have occurred at higher momenta in the Sun's past when solar flares were more powerful. For instance, \citet{atri_2017} uses a large SEP event as a representative spectrum for an M dwarf star which has a higher cut-off energy at approximately $\sim$GeV energies.

\begin{figure}
	\centering
        \includegraphics[width=0.5\textwidth]{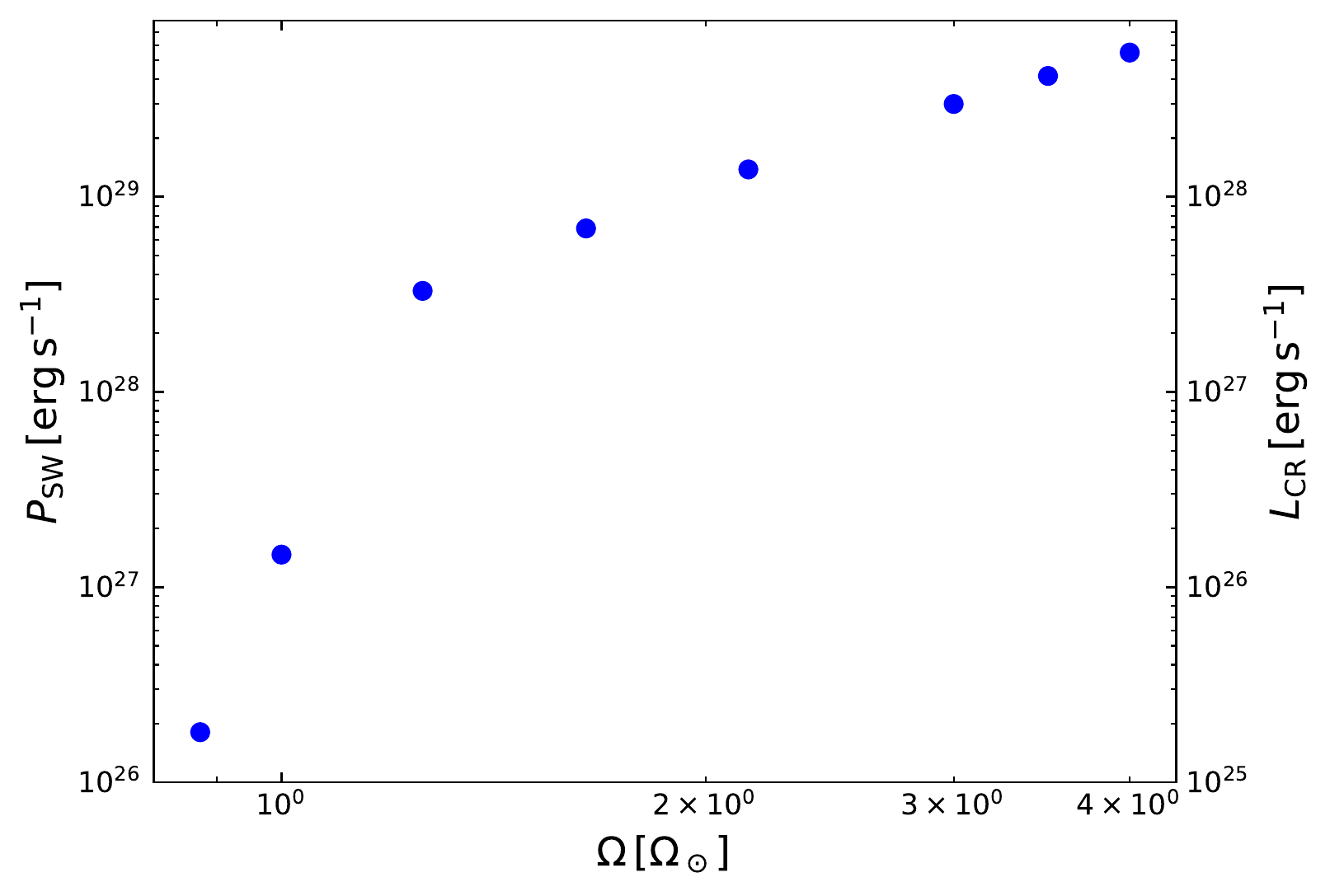}
       	\centering
  \caption{Kinetic power in the solar wind, $P_\mathrm{SW}$, as a function of stellar rotation rate. The righthand side of the plot indicates the total kinetic power that we inject in stellar cosmic rays corresponding to 10\% of $P_\mathrm{SW}$. } 
    \label{fig:omega_psol}
\end{figure}

\subsubsection{Total kinetic power in stellar cosmic rays}
\label{subsubsec:pcr}
We use the kinetic power in the stellar wind, $P_\mathrm{SW}=\dot M(\Omega)v_\infty(\Omega)^2/2$, calculated from our stellar wind model to estimate $L_\mathrm{CR}$ as a function of stellar rotation rate assuming a certain efficiency, shown in Fig.\,\ref{fig:omega_psol}. $\dot M(\Omega)$ and $v_\infty(\Omega)$ are the mass loss rate and terminal speed of the stellar wind, respectively. Here we assume that $L_\mathrm{CR}\sim 0.1P_\mathrm{SW}$ which is shown on the righthand side of Fig.\,\ref{fig:omega_psol}. Such a value is typical of the equivalent efficiency factor estimated for supernova remnants \citep[][for instance]{vink_2010}. Without further evidence to go by, we simply adopt the same value here for young stellar flares. Adopting a different efficiency of 1\% or 100\%, for instance, would change the values for the differential intensity of stellar cosmic rays presented in Section\,\ref{sec:results} by a factor of 0.1 and 10, respectively. The values that we use here are broadly in line with the value of $L_\mathrm{CR} \sim 10^{28}\mathrm{erg\,s^{-1}}$ from \citet{rodgers-lee_2017} which was motivated as the kinetic power of stellar cosmic rays produced by a T-Tauri star.

\subsection{Comparison of solar, stellar and Galactic cosmic ray spectra}
\label{subsubsec:sketch}
\begin{figure*}
	\centering
        \includegraphics[width=0.7\linewidth]{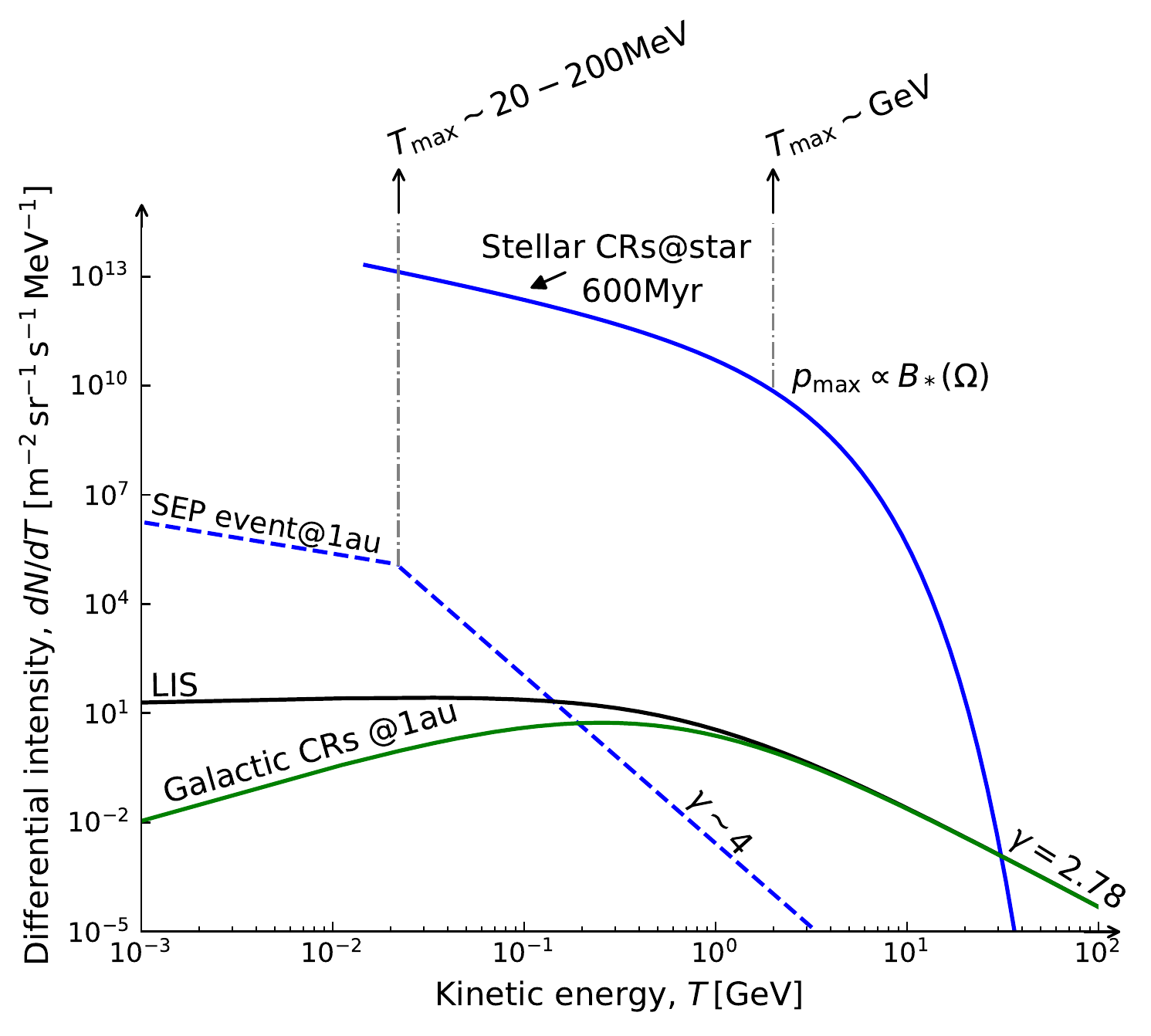}
       	\centering
  \caption{This sketch illustrates typical cosmic ray spectra, both solar/stellar (impulsive and gradual events) and Galactic in origin, at various orbital distances in the solar/stellar system. The solid black line represents an approximation for the LIS of Galactic cosmic rays outside of the solar system. The green line represents a typical Galactic cosmic ray spectrum observed at Earth. The blue dashed line is a typical spectrum for a gradual SEP event (averaged over the duration of the event). The spectral slope $\gamma$ at high energies is also indicated in the plot. The solid blue line is the estimate for an impulsive stellar energetic particle event spectrum that we motivate in this paper for a young solar-type star ($\sim 600$Myr old) which includes a spectral break at much higher energies than the typical present-day (gradual) SEP spectrum (blue dashed line). See Section\,\ref{subsubsec:sketch} for more details. } 
    \label{fig:sketch}
\end{figure*}

We include a schematic in Fig.\,\ref{fig:sketch} which shows representative values for the differential intensity of solar and Galactic cosmic rays as a function of kinetic energy. The differential intensity of cosmic rays, $j$, is often considered (rather than the phase space density given in Eq.\,\ref{eq:f}) as a function of kinetic energy which we plot in Section\,\ref{sec:results}. These quantities are related by $j(T)=dN/dT=p^2f(p)$. 

Fig.\,\ref{fig:sketch} includes the estimate for our most extreme steady-state spectrum for stellar cosmic rays injected close to the stellar surface for $\Omega=4\Omega_\odot$ or $t_*=600\,$Myr (solid blue line). This is representative of an impulsive stellar energetic particle event. The spectral break occurs at $\sim$GeV energies as motivated in the previous section. The resulting spectrum at 1\,au (and other orbital distances) is presented in Section\,\ref{sec:results}. 

\begin{table*}
\centering
\caption{List of parameters for the simulations. The columns are, respectively: the age ($t_*$) of the solar-type star, its rotation rate ($\Omega$) in terms of the present-day solar value ($\Omega_\odot = 2.67\times 10^{-6}\,\mathrm{rad\,s^{-1}}$), its rotation period ($P_\mathrm{rot}$), the astrospheric radius ($R_\mathrm{h}$), the radial velocity ($v_\mathrm{1au}$) and the magnitude of the total magnetic field ($|B_\mathrm{1au}|$) at $r=1$\,au. $\dot M$ is the mass-loss rate. $L_\mathrm{CR}$ is the power we inject in stellar cosmic rays which we relate to the kinetic power in the stellar wind. The second and third last columns are the momentum and kinetic energy for the stellar cosmic rays at which the exponential break in the injected spectrum occurs. In order to reproduce our injected cosmic ray spectrum ($Q$ in Eq.\,\ref{eq:f}-\ref{eq:q}): first, the values of $L_\mathrm{CR}$ and $p_\text{max}$ given below can be used in Eq.\,\ref{eq:rateb} to determine $(\frac{d\dot N}{dp})|_{2mc}$ for a given value of $\Omega$. Then, $\frac{d\dot N}{dp}$, and therefore $Q$, can be calculated for a given value of $\Omega$.
The last column gives the kinetic energy below which the stellar cosmic ray intensities dominate over the Galactic cosmic ray intensities at 1\,au.}
\begin{tabular}{@{}ccccccccccccccc@{}}
\hline

$t_*$ &$\Omega$ &$P_\mathrm{rot}$& $R_\mathrm{h}$ & $v_\mathrm{1au}$ &$|B_\mathrm{1au}|$ & $\dot M$ &   $L_\mathrm{CR}=0.1P_\mathrm{SW}$ & $p_\mathrm{max}$ & $T_\mathrm{max}$ & $T(j^\text{SCR}_\text{1\,au}=j^\text{GCR}_\text{1\,au})$\\
\hline
[Gyr] & $[\Omega_\odot]$&[days]   &[au] & $\mathrm{[km\,s^{-1}]}$ & [G] &[$M_\odot\,\mathrm{yr}^{-1}$]  & $[\mathrm{erg\,s^{-1}}$] & $[\mathrm{GeV}/c]$ & $[\mathrm{GeV}]$ & [GeV]\\
\hline
2.9 			 & 1.3 	 & 22   & 500  & 610 &$5.4\times 10^{-5}$ & $2.8\times 10^{-13}$ & $3.30\times10^{27}$ & 0.26 & 0.04 & 1.3 \\
1.7 			 & 1.6 	 & 17   & 696  & 660 &$7.6\times 10^{-5}$ & $5.1\times 10^{-13}$ &  $6.89\times10^{27}$ & 0.38 &0.07 & 2.6 \\ 
1.0 			 & 2.1   & 13   & 950  & 720 &$1.2\times 10^{-4}$ & $8.5\times 10^{-13}$ & $1.38\times10^{28}$ & 0.54 &0.14  & 4.1 \\ 
0.6 			 & 3.0	 & 9    & 1324 & 790 &$1.8\times 10^{-4}$ & $1.5\times 10^{-12}$ &  $2.99\times10^{28}$ & 0.85 &0.33 & 8.0 \\  
0.6 			 & 3.5   & 8    & 1530 & 820 &$2.8\times 10^{-4}$ & $2.0\times 10^{-12}$ &  $4.16\times10^{28}$ & 1.03 &0.46 & 10 \\ 
0.6 			 & 4.0 	 & 7    & 1725 & 850 &$3.5\times 10^{-4}$ & $2.4\times 10^{-12}$ & $5.49\times10^{28}$  & 1.23 &0.61 & 13 \\ 
\hline 
0.6 			 & 3.5   & 8    & 1530 & 820 &$2.8\times 10^{-4}$ & $2.0\times 10^{-12}$ &  $4.16\times10^{28}$ & 3.30 &2.49 & 33 \\
\hline \\
\end{tabular}

\label{table:sim_parameters}
\end{table*}

A fit to the Galactic cosmic ray LIS, constrained by the {\it Voyager\,1} observations \citep{stone_2013,cummings_2016,stone_2019} outside of the heliosphere, is denoted by the solid black line in Fig.\,\ref{fig:sketch} \citep[Eq.\,1 from][]{vos_2015}. A fit to the modulated Galactic cosmic ray spectrum measured at Earth is given by the solid green line \citep[using the modified force field approximation given in Eq.10 of][with $\phi=0.09\,$GeV]{rodgers-lee_2020b}. We also indicate the spectral slope, $\gamma=2.78$, at high energies on the plot. Note, this represents $dN/dT \propto T^{-\gamma}$ rather than $dN/dp \propto p^{-\alpha}$. The power law indices $\gamma$ and $\alpha$ are related. At relativistic energies, $\gamma = \alpha$ since $T =pc$ and at non-relativistic energies $\alpha = 2\gamma-1$.
 It is important to note that the measurements at Earth change a certain amount as a function of the solar cycle. Here, however we treat the LIS and the Galactic cosmic ray spectrum at Earth as constant when making comparisons with the stellar cosmic ray spectrum as a function of stellar rotation rate\footnote{
It is important to note that the Galactic cosmic ray LIS may have been different in the past. Supernova remnants are believed to be a major contributor to the Galactic component of the LIS \citep{drury_1983,drury_2012}. The star formation rate (SFR, which can be linked to the number of supernova remnants using an initial mass function) of the Milky Way in the past therefore should influence the LIS in the past. For instance, high ionisation rates (with large uncertainties) have been inferred for galaxies at high redshifts which have higher SFRs than the present-day Milky Way \citep{muller_2016, indriolo_2018}. In these studies, the inferred ionisation rate is attributed to galactic cosmic rays. Recently, using observations of the white dwarf population in the solar neighbourhood ($d<100$pc), \citet{isern_2019} reconstructed an effective SFR for the Milky Way in the past. They found evidence of a peak in star formation $\sim 2.2-2.8$\,Gyr ago, an increase by a factor of $\sim$3 in comparison to the present-day SFR. Using a sample of late-type stars, \citet{rocha-pinto_2000} also found an increase in the SFR by a factor of $\sim$2.5 approximately $2-2.5$\,Gyr ago. Since these results suggest that the SFR has been within a factor of $\sim$3 of its present-day value for the stellar ages that we focus on, we did not vary the LIS fluxes with stellar age in \citet{rodgers-lee_2020b}.}. 

On the other hand, the differential intensities for SEPs observed at Earth cannot be treated as constant in time. The SEP spectrum at 1\,au, shown by the blue dashed line in Fig.\,\ref{fig:sketch}, is not continuous in time for the present-day Sun. The differential intensity given by the dashed blue line represents the typical intensities of SEPs at Earth that are derived from time-averaged observations of particle fluences \citep[such as those presented in][]{mewaldt_2005}. This spectrum is representative of a gradual SEP event. This type of SEP event lasts approximately a few days. 

\citet{rab_2017} estimated the stellar cosmic ray spectrum for a young pre-main sequence star (shown in their Fig.\,2) representing the present-day values for a typical gradual SEP event multiplied by a factor of $10^5$ \citep[the motivation for which is given in][]{feigelson_2002}. \citet{tabataba-vakili_2016} similarly use a typical spectrum for a gradual solar energetic particle event and scale it with $1/R^2$ to 0.153\,au in order to find the values for the differential intensity of stellar cosmic rays at the location of a close-in exoplanet orbiting an M dwarf star. In both of these examples the spectral shape is held constant, whereas here it is not. The propagation of stellar cosmic rays from the star/CME through the stellar system is not the focus of either of these papers. This type of treatment for estimating the spectrum of stellar cosmic ray events at 1\,au, or other orbital distances, around younger stars (and later type stars) and the impact of the underlying assumptions are what we investigate in this paper. This can be used as a starting point towards deriving more realistic stellar cosmic ray spectra in the future that can be constrained by upcoming missions like JWST and Ariel \citep{tinetti_2018}. 

Transmission spectroscopy using JWST will be able to detect emission features from molecules in exoplanetary atmospheres. Stellar and Galactic cosmic rays should produce the same chemical reactions. Thus, close-in exoplanets around young and/or active stars are the best candidates to detect the chemical signatures of stellar cosmic rays as they should be exposed to high stellar cosmic ray fluxes. In comparison, the Galactic cosmic ray fluxes at these orbital distances should be negligible. \citet{helling_2019} and \citet{barth_2020} identify a number of ``fingerprint ions" whose emission, if detected in an exoplanetary atmosphere, would be indicative of ionisation by cosmic rays. These fingerprint ions are ammonium ($\mathrm{NH_4^+}$) and oxonium ($\mathrm{H_3O^+}$). \citet{barth_2020} also suggest that stellar and Galactic cosmic rays contribute (along with other forms of high energy radiation, such as X-rays) to the abundance of the following key organic molecules: hydrogen cyanide (HCN), formaldehyde ($\mathrm{CH_2O}$) and ethylene ($\mathrm{C_2H_4}$). \citet{barth_2020} indicate that $\mathrm{CH_2O}$ and $\mathrm{C_2H_4}$ may be abundant enough to possibly be detected by JWST.


\subsection{Overview of the simulations}
We consider 7 cosmic ray transport simulations in total for our results. Additional test case simulations are presented in Appendix\,\ref{sec:test} for physical set-ups with known analytic solutions verifying that our numerical method reproduces well these expected results. Six of the 7 simulations that we ran represent the result of varying the stellar rotation rate. The remaining simulation, for $\Omega = 3.5\Omega_\odot$, investigates the effect of increasing the value of the $p_\mathrm{max}$ which is discussed in Appendix\,\ref{subsec:pb}. The parameters for the simulations are shown in Table\,\ref{table:sim_parameters}.

The values for the astrospheric radii, $R_\mathrm{h}(\Omega)$, are given in Table\,\ref{table:sim_parameters} which is the outer radial boundary. These values were derived by balancing the stellar wind ram pressure against the ram pressure of the ISM \citep[see Section 2.3.3 of][]{rodgers-lee_2020b}. The logarithmically spaced radial bins for $i=0,...,N$ are given by $r_i = \mathrm{exp}\{i\times\mathrm{ln}(r_N/r_0)/(N-1) + \mathrm{ln}\,r_0\}$ where $r_0=1\,R_\odot$ and $r_N=R_\mathrm{h}(\Omega)$ with $N=60$.

\section{Results}
\label{sec:results}

In this section we investigate the evolution of the stellar cosmic ray spectrum at different orbital distances as a function of stellar rotation rate. Five parameters vary with $\Omega$ for these simulations: $B(r)$, $v(r)$, $R_\text{h}$, $L_\mathrm{CR}$ and $p_\mathrm{max}$. The value of $R_\text{h}$ does not play much of a role in our simulations since it is always much larger than the orbital distances that we are interested in.

After travelling through the stellar wind, stellar cosmic rays can interact with a planet's atmosphere. If a planetary magnetic field is present this will also influence the propagation of the stellar cosmic rays through the atmosphere \citep[e.g.][]{griessmeier_2015}. Higher energy cosmic rays will be less easily deflected by an exoplanetary magnetic field. Cosmic rays with energies that are capable of reaching the surface of an exoplanet are of interest for the origin of life. For this, the pion production threshold energy of 290\,MeV should be significant. Pions produce secondary particles which can trigger particle showers \citep[as discussed in][for instance]{atri_2020}. Sufficiently energetic secondary particles, such as neutrons, can reach the surface of a planet which are known as ground level enhancements. Solar neutrons have been detected even on Earth with neutron monitors since the 1950s \citep{simpson_1951}. Thus, our aim is to determine the range of stellar rotation rates for which the differential intensity of stellar cosmic rays dominates over Galactic cosmic rays at energies above the pion threshold energy. 

\begin{figure*}
        \includegraphics[width=\textwidth]{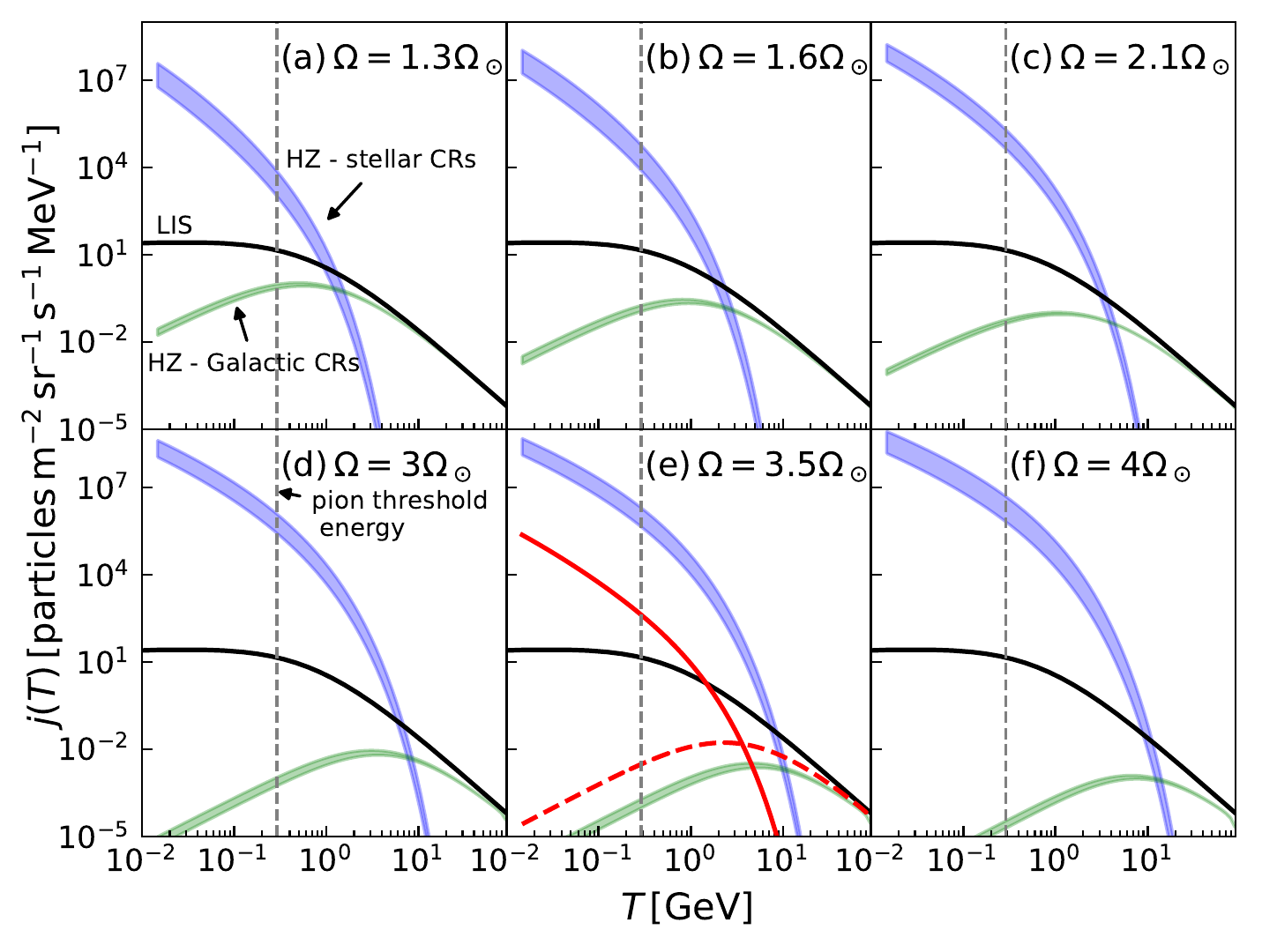}
       	\centering
    \caption{The differential intensity of stellar cosmic rays (blue shaded regions) and Galactic cosmic rays (green shaded regions) in the habitable zone as a function of kinetic energy. Each panel represents a different value for the stellar rotation rate. $\Omega=2.1\Omega_\odot$ corresponds to $t_*=1$\,Gyr, shown in (c), when life is thought to have begun on Earth. Also shown are the differential intensities of stellar (solid red line) and Galactic cosmic rays (red dashed line) at 20\,au, the orbital distance of HR 2562b, in panel (e). The black solid line is a fit to the \emph{Voyager 1} data for the LIS. The grey dashed line represents the pion threshold energy, 290\,MeV. See text in Section\,\,\ref{subsec:scr_omega}.  }
\label{fig:omega}
\end{figure*}

\subsection{Stellar cosmic rays as a function of stellar rotation rate (or age)}
\label{subsec:scr_omega}
Fig.\,\ref{fig:omega} shows the stellar cosmic ray differential intensities as a function of kinetic energy for our simulations. In each of the panels the blue shaded region represents the values of differential intensities for stellar cosmic rays present in the habitable zone for a solar-mass star. For comparison, the green shaded region shows the differential intensities for Galactic cosmic rays in the habitable zone \citep[from the simulations presented in][]{rodgers-lee_2020b}.

The habitable zone of a solar-mass star evolves with stellar age which we have incorporated in the shaded regions of Fig.\,\ref{fig:omega}. We follow the formalism of \citet{selsis_2007} with the recent Venus and early Mars criteria, using the stellar evolutionary model of \citet{baraffe_1998}. At 600\,Myr the young Sun was less luminous and had an effective temperature slightly smaller than its present day value. Thus, the habitable zone at 600\,Myr was located closer to the Sun between $r \sim 0.64-1.58$\,au in comparison to the present day values of $r\sim 0.72-1.77$\,au (using the recent Venus and early Mars criteria). Given the finite resolution of our spatial grid some of the blue shaded regions in Fig.\,\ref{fig:omega} are slightly smaller than the calculated habitable zone. Finally, for comparison in each of the panels the solid black line shows the LIS values \citep[from][]{vos_2015}. The vertical grey dashed line represents the pion threshold energy at 290\,MeV.

Figs.\,\ref{fig:omega}(a)-(f) show at the pion threshold energy that stellar cosmic rays dominate over Galactic cosmic rays in the habitable zone for all values of stellar rotation rate (or age) that we consider. The energy that they dominate up to differs though as a function of stellar rotation rate (given in Table\,\ref{table:sim_parameters}). For instance, at $\Omega=1.3\,\Omega_\odot$, the transition from stellar cosmic rays dominating over Galactic cosmic rays occurs at $\sim1.3\,$GeV. It increases up to $\sim13\,$GeV for $\Omega=4\,\Omega_\odot$. The stellar cosmic ray fluxes also increase in the habitable zone as a function of stellar rotation rate. At the same time, the Galactic cosmic ray fluxes decrease. 

The red dashed line and solid lines in Fig.\,\ref{fig:omega}(e) are the values for the differential intensities at 20\,au for Galactic and stellar cosmic rays, respectively. We previously discussed the Galactic cosmic ray differential intensities for $\Omega = 3.5\Omega_\odot$ (Fig.\,\ref{fig:omega}(e) here) in \citet{rodgers-lee_2020b} in the context of the HR2562 exoplanetary system. HR2562 is a young solar-like star with a warm Jupiter exoplanet orbiting at 20\,au. Although Galactic cosmic rays (dashed red line) represent a source of continuous cosmic ray flux, stellar cosmic rays can dominate (solid red line) at approximately the orbital distance of the exoplanet for $\lesssim$ 5\,GeV. This would happen at times of impulsive events.

The solid blue line in Fig.\,\ref{fig:sketch} shows the steady-state spectrum close to the star corresponding to $\Omega = 4\Omega_\odot$. By comparing with the values for the fluxes found in the habitable zone, shown in Fig.\,\ref{fig:omega}(f), we can determine by how many orders of magnitude the stellar cosmic ray fluxes have decreased between $\sim1\,R_\odot$ and $\sim$1\,au ($\sim200\,R_\odot$). The decrease is slightly greater than 4 orders of magnitude. The decrease is the combined result of diffusive and advective processes. In Appendix\,\ref{sec:test}, we discuss the effect of the different physical processes, shown in Fig.\,\ref{fig:appendix}. 

\begin{figure}
	\centering
        \includegraphics[width=0.5\textwidth]{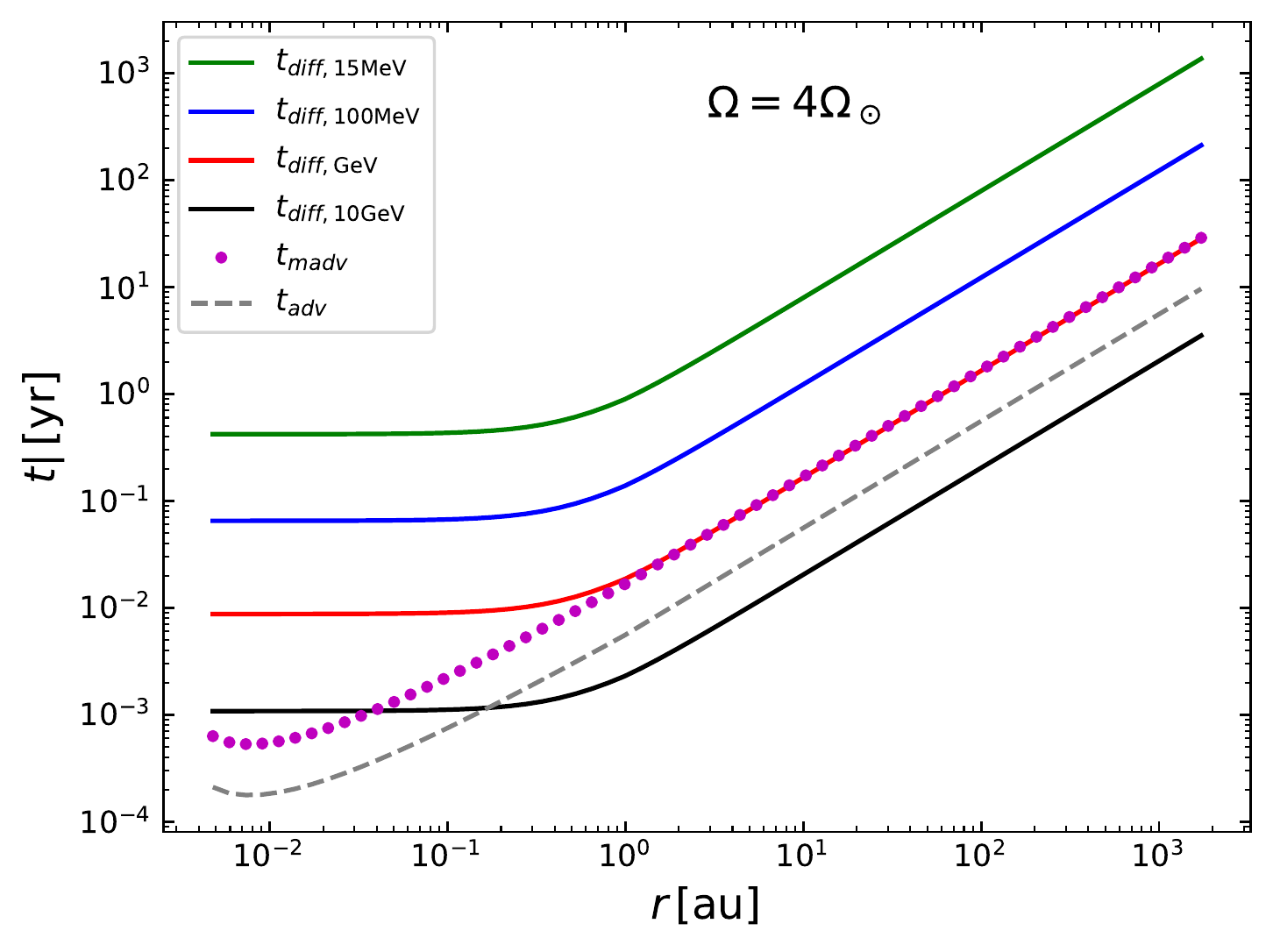}
       	\centering
  \caption{Timescales for the different physical processes for the stellar wind properties corresponding to a stellar rotation rate of $\Omega = 4\Omega_\odot$, corresponding to $t_*\sim 600\,$Myr. The solid lines represent the diffusion timescale for cosmic rays with different energies. The magenta dotted line and the grey dashed line represent the momentum advection and advection timescales, respectively. For 10\,GeV cosmic rays, $t_\mathrm{madv}\lesssim t_\mathrm{diff}$ at $r\lesssim 0.03\,$au and $t_\mathrm{madv}\lesssim t_\mathrm{diff}$ at $r\lesssim 0.5\,$au for GeV energies. This illustrates the importance of adiabatic losses for the stellar cosmic rays at small orbital distances.  } 
    \label{fig:timescales_omega21}
\end{figure}

Fig.\,\ref{fig:timescales_omega21} shows the timescales for the different physical processes for $\Omega = 4\Omega_\odot$. The diffusion timescales for 0.015, 0.1, 1 and 10 GeV energy cosmic rays are shown by the solid lines in Fig.\,\ref{fig:timescales_omega21} where $t_\mathrm{diff} = r^2/\kappa(r,p,\Omega)$. The magenta dots represent an estimate for the momentum advection timescale $t_\mathrm{madv}\sim 3r/v$. For $r\lesssim$1\,au, Fig.\,\ref{fig:timescales_omega21} shows that the spatial and momentum advection timescales are shorter than the diffusion timescale for cosmic rays with kinetic energies $\lesssim$GeV. These low energy cosmic rays are affected by adiabatic losses in this region and are being advected by the stellar wind, rather than propagating diffusively. Since the stellar cosmic rays are injected close to the surface of the star only the cosmic rays with kinetic energies $\gtrsim$GeV, and therefore short diffusion timescales, propagate diffusively out of this region.

We also investigated the sensitivity of our results on our choice of $p_{\rm max}$ in Appendix\,\ref{subsec:pb}. Fig.\,\ref{fig:comparisons} shows the results of adopting a higher maximum cosmic ray momentum for $\Omega = 3.5\Omega_\odot$. We find that the location of the stellar cosmic ray spectral break is an important parameter to constrain and that it affects our results significantly, with the maximum energy at which stellar cosmic rays dominate Galactic cosmic rays being an increasing function of $p_\mathrm{max}$.

\begin{figure}
        \includegraphics[width=\columnwidth]{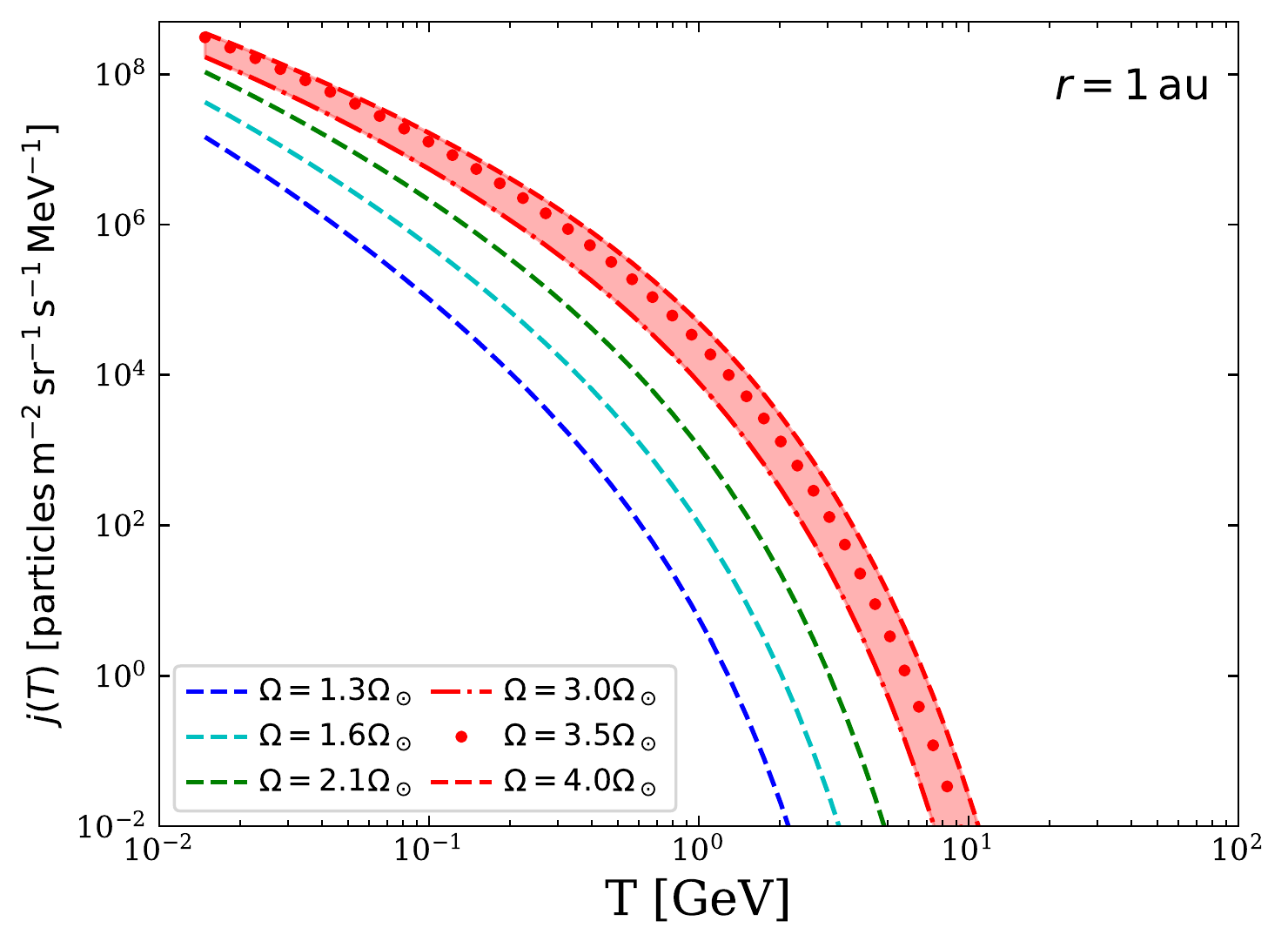}
       	\centering
    \caption{Plot of the differential intensities for different stellar rotation rates at 1\,au. The shaded red region represents different stellar rotation rates corresponding to the same stellar age, namely 600\,Myr. }
\label{fig:junnormalised}
\end{figure}

\subsection{Differential intensities as a function of rotation rate at 1\,au}
Fig.\,\ref{fig:junnormalised} shows the differential intensities of the stellar cosmic rays at 1\,au as a function of $\Omega$. The differential intensities obtained at 1\,au increase as a function of stellar rotation rate. The increase in the differential intensities is almost entirely due to the corresponding increase in $L_\text{CR}$. The red shaded region indicates the values for different stellar rotation rates with the same stellar age, $t_*=600\,$Myr. The shift in the maximum energy to higher energies with increasing stellar rotation rate can also been seen by comparing the $\Omega=1.3\Omega_\odot$ (dashed blue line) and $\Omega=4\Omega_\odot$ (dashed red line) cases. The slope of the spectrum at $10^{-2}-1$\,GeV energies becomes less steep with increasing stellar rotation and starts to turn over at slightly higher energies.

\subsection{Assumption of continuous injection}
\label{subsec:injection_assumption}
Fig.\,\ref{fig:omega} shows that the intensities of stellar cosmic rays are greater than those of the Galactic cosmic rays at energies around the pion energy threshold for all values of stellar rotation rate that we consider. However, we must also estimate the energy up to which these stellar cosmic rays can be treated as continuous in time. 

In order for the stellar cosmic ray flux to be considered continuous, the rate of flare events (producing the stellar cosmic rays) must be larger than the transport rate for a given cosmic ray energy. We use $1/t_\mathrm{diff}$ at $\lesssim 1\,$au where it is approximately independent of radius as a reference value for the transport rate. We estimate the maximum stellar cosmic ray energy which can be taken as continuous by considering the relation between flare energy and flare frequency \citep[$dN/dE_\mathrm{flare}\propto E_\mathrm{flare}^{-1.8}$ from][]{maehara_2015}. First, from Fig.\,4 of \citet{maehara_2015} we can obtain the flare rate, by multiplying the flare frequency by the flare energy, as a function of flare energy. Fig.\,2 of \citet{maehara_2015} also indicates that stars with rotation periods between 5-10\,days flare approximately 10 times more frequently than slow rotators, like the Sun. Thus, as an estimate we increase the flare rate by an order of magnitude for fast rotators as a function of flare energy \citep[Fig.\,4 of][]{maehara_2015}. We determine $p_\mathrm{max}$ for a given flare energy by equating the flare energy with magnetic energy such that $E_\text{flare}\propto B^2$ \citep[similar to][]{herbst_2020b}. Therefore, using the Hillas criterion given in Eq.\,\ref{eq:hillas}, $p_\mathrm{max} \propto E_\text{flare}^{1/2}$.

\begin{figure}
        \includegraphics[width=\columnwidth]{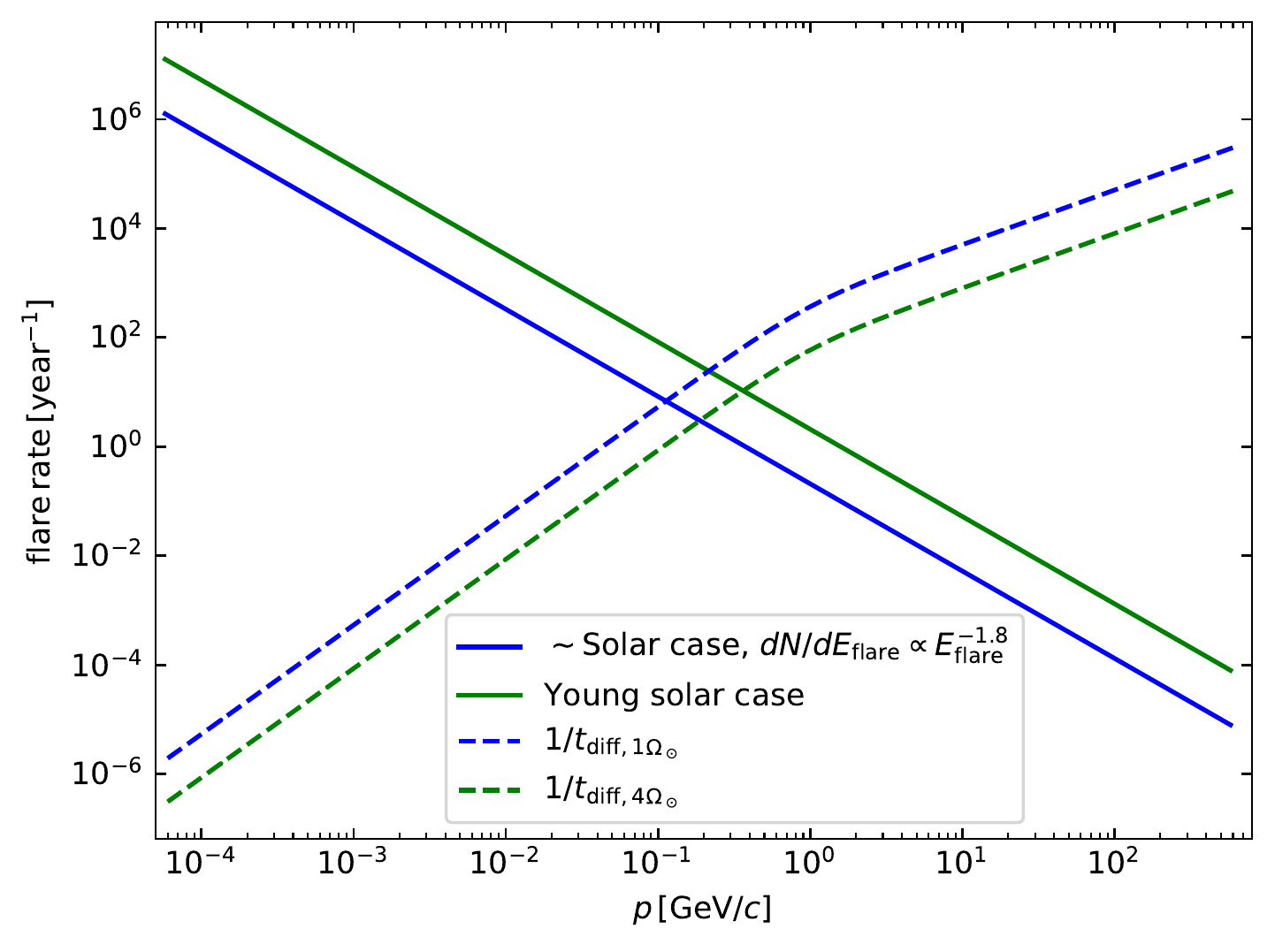}
       	\centering
    \caption{Plot of flare rate \citep[flare frequency from][times $E_\mathrm{flare}$]{maehara_2015} as a function of $p_\mathrm{c,max}$ for a slow rotator like the Sun (with $P_\mathrm{rot}>20\,$days, solid blue line) and for fast rotators (with $P_\mathrm{rot}=5-10\,$days, solid green line) at 1\,au. The diffusive transport rate as a function of momentum is overplotted for $\Omega = 1\Omega_\odot$ and $\Omega = 4\Omega_\odot$ by the dashed blue and green lines, respectively. }
\label{fig:flare_rate}
\end{figure}

In Fig.\,\ref{fig:flare_rate}, we plot the flare rate (solid lines) and diffusion rates (dashed lines) as a function of momentum. The diffusive timescale for the slow rotator/$\sim$solar case is based on the stellar wind properties presented in \citet{rodgers-lee_2020b} for the present-day Sun, $\Omega=1\Omega_\odot$. For the slow rotator/solar case, this plot indicates that the maximum continuously injected cosmic ray momentum is $p_\mathrm{c,max}=0.11$\,GeV$/c$ ($T_\mathrm{c,max}=5\,$MeV). For fast rotators, it indicates that $p_\mathrm{c,max}=0.4$\,GeV$/c$ ($T_\mathrm{max}=80\,$MeV). Thus, even for our most extreme case, flare-injected stellar cosmic rays cannot be considered as continuous beyond 80\,MeV in energy. The plot has been normalised such that $\sim$GeV cosmic ray energies correspond to $E_\mathrm{flare}\sim10^{33}$erg. It is important to note that here we have determined quite low values of $p_\mathrm{c,max}$ by comparing the diffusive transport rate with the flare rate. However, a comparison of the flare rate with the chemical recombination rates in exoplanetary atmospheres may result in higher values for $p_\mathrm{c,max}$. 

\section{Discussion \& conclusions}
\label{sec:conclusions}
In this paper we have investigated the differential intensity of stellar cosmic rays that reach the habitable zone of a solar-type star as a function of stellar rotation rate (or age). We motivated a new spectral shape for stellar cosmic rays that evolves as a function of stellar rotation rate. In particular, the maximum injected stellar cosmic ray energy and total injected stellar cosmic ray power evolve as a function of stellar rotation rate. We consider stellar cosmic rays injected at the surface of the star, which would be associated with stellar flares whose solar counterpart are known as impulsive SEP events. The values for the total injected stellar cosmic ray power and the maximum stellar cosmic ray energy that we provide in this paper can be used to reproduce our injected stellar cosmic ray spectrum. We then used the results of a 1.5D stellar wind model for the stellar wind properties \citep[from][]{rodgers-lee_2020b} in combination with a 1D cosmic ray transport model to calculate the differential intensity of stellar cosmic rays at different orbital distances.

Our main findings are that, close to the pion threshold energy, stellar cosmic rays dominate over Galactic cosmic rays at Earth's orbit for the stellar ages that we considered, $t_*=0.6-2.9$\,Gyr ($\Omega =1.3-4\Omega_\odot$). Stellar cosmic rays dominate over Galactic cosmic rays up to $\sim10\,$GeV energies for stellar rotation rates $>3 \Omega_\odot$, corresponding approximately to a stellar age of 600\,Myr. The differential intensities of the stellar cosmic rays increases with stellar rotation rate, almost entirely due to the increasing stellar cosmic ray luminosity. At 1\,Gyr, when life is thought to have begun on Earth, we find that high fluxes of stellar cosmic rays dominate over Galactic cosmic rays up to 4\,GeV energies. However, based on stellar flare rates, we estimate that the stellar cosmic ray fluxes may only be continuous in time up to MeV energies even for the fastest rotator cases that we consider. For momenta where the diffusive transport rate is larger than the flare rate, the flare injection cannot be treated as continuous. The transition point corresponds to $p_\mathrm{c,max} = 0.1$ and $0.4$\,GeV$/c$, or to $T_\mathrm{c,max} = 5$ and 80\,MeV, for the slow rotator/solar and young solar cases, respectively.

Our results overall highlight the importance of considering stellar cosmic rays in the future for characterising the atmospheres of young close-in exoplanets. They also highlight the possible importance of stellar cosmic rays for the beginning of life on the young Earth and potentially on other exoplanets.

We find for the young exoplanet HR\,2562b, orbiting its host star at 20\,au, that stellar cosmic rays dominate over Galactic cosmic rays up to $\sim4\,$GeV energies despite the large orbital distance of the exoplanet. However, these stellar cosmic ray fluxes may not be continuous in time.

Our results presented in Fig.\,\ref{fig:omega}, for a stellar age of 600\,Myr ($\Omega=4\Omega_\odot$), demonstrate that low energy stellar cosmic rays ($<$GeV) move advectively as they travel out through the stellar wind from the injection region to 1\,au. In this region the low energy cosmic rays are also impacted by adiabatic losses. Beyond 1\,au the low energy cosmic rays are influenced to a greater extent by diffusion. 
This finding is quite interesting because the velocity of the solar wind at close distances is currently unknown. NASA's Parker Solar Probe \citep{bale_2019} and ESA's Solar Orbiter \citep{owen_2020} have only recently begun to probe the solar wind at these distances. Thus, our simulation results are sensitive to parameters of the solar wind that are only now being observationally constrained. If the solar wind is faster in this region than what we have used in our models then the fluxes of stellar cosmic rays that we calculate at larger radii will be smaller.

Our results are based on a 1D cosmic ray transport model coupled with a 1.5D stellar wind model. In reality, stellar winds are not spherically symmetric. Latitudinal variations are seen in the solar wind which also depend on the solar cycle \citep[e.g.][]{mccomas_2003} and magnetic maps of other low-mass stars also show that the magnetic field structure is not azimuthally symmetric \citep[e.g.][]{llama_2013,vidotto_2014b}. Gradients in the magnetic field can lead to particle drifts which we cannot investigate with our models. Our results are based on steady-state simulations which means that effects occurring on timescales shorter than the rotation period of the star are neglected in the cosmic ray transport model. The fact that flares may also occur at positions on the stellar surface which then do not reach Earth is not taken into account in our models. It will be of great interest in the future to use 2D or 3D cosmic ray transport models in combination with 3D stellar wind models \citep[e.g.][]{kavanagh_2019,folsom_2020} to study in greater detail the stellar cosmic ray fluxes reaching known exoplanets. Our results represent some type of average behaviour that could be expected: at particular times during a stellar cycle the stellar cosmic ray production rate via flares could be increased, whereas at other times during the minimum of a stellar cycle the production rate would be lower. However, due to the present lack of observational constraints for the stellar cosmic ray fluxes in other stellar systems using a simple 1D cosmic ray transport model and a 1.5D stellar wind model is justified.

Finally, it is also worth bearing in mind that the stellar cosmic rays considered here are representative of impulsive events. The stellar cosmic ray fluxes produced by CMEs are likely to be far in excess of those presented here. These fluxes would be even more transient in nature than the stellar cosmic ray fluxes presented here. In light of these findings, future modelling of stellar cosmic rays from transient flare events and gradual events appears motivated.

\section*{Acknowledgements}
DRL and AAV acknowledge funding from the European Research Council (ERC) under the European Union's Horizon 2020 research and innovation programme (grant agreement No 817540, ASTROFLOW). The authors wish to acknowledge the DJEI/DES/SFI/HEA Irish Centre for High-End Computing (ICHEC) for the provision of computational facilities and support. DRL would like to thank Christiane Helling for very helpful discussions which improved the paper. We thank the anonymous reviewer for their constructive comments.

\appendix
\section{Test cases}
\label{sec:test}
We present three simulations to illustrate that the code reproduces the expected analytic results for a number of simple test cases. We isolate the effect of different physical terms in Eq.\,\ref{eq:f}, giving additional insight into the system. The test cases use the stellar wind parameters for the $\Omega = 3.5\Omega_\odot$ simulation unless explicitly stated otherwise. For all of the test simulations the same power law is injected as described in Eq.\,\ref{eq:rateb} with $p_\mathrm{max} = 1.03\,\mathrm{GeV/}c$ and $L_\mathrm{CR}=4.16\times10^{28}\,\mathrm{erg\,s^{-1}}$.

\begin{figure}
	\centering
        \includegraphics[width=0.5\textwidth]{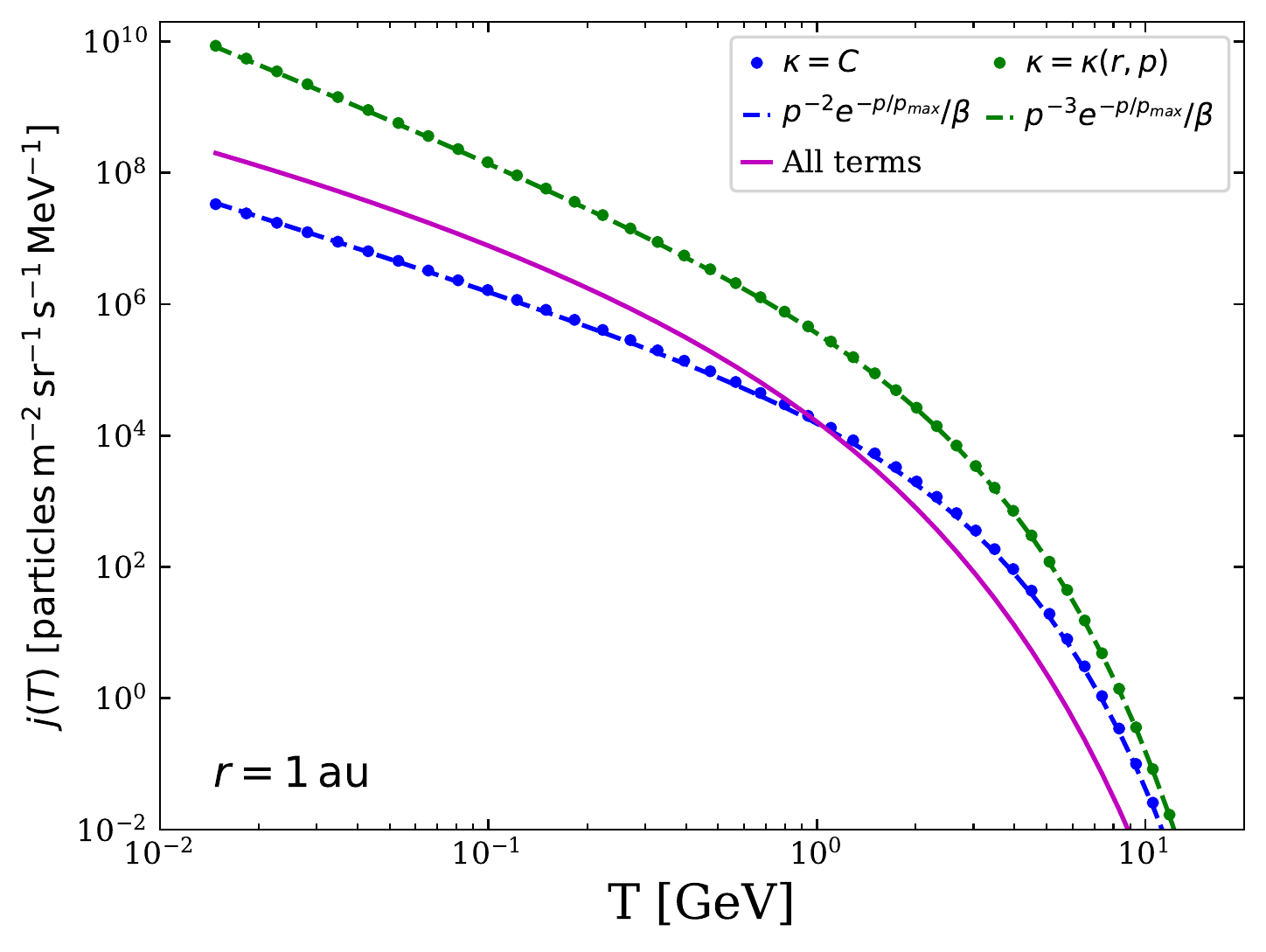}
       	\centering
  \caption{The differential intensity for stellar cosmic rays as a function of kinetic energy at 1\,au are shown here for a number of test cases, described in Section\,\ref{sec:test}.  The blue dots are the values obtained using a constant diffusion coefficient, test case (a). The green dots represent the model with only spatial diffusion, test case (b). Finally, the magenta solid line includes all terms considered in our models, test case (c). Different power laws are shown by the dashed lines. \label{fig:appendix}} 
\end{figure}

The three test cases are simulations with: (a) a constant diffusion coefficient only, (b) the momentum-dependent diffusion coefficient derived from the magnetic field profile for $\Omega = 3.5\Omega_\odot$ only and (c) the diffusion coefficient used for (b) along with the spatial and momentum advection terms. These test cases are described below in more detail. The results from these tests are shown in Fig.\,\ref{fig:appendix} for $1\,$au.

The first test case consisted of using a constant diffusion coefficient in momentum and space with $-v\cdot\nabla f=((\nabla \cdot v)/3) (\partial f/\partial\mathrm{ln}p)=0$ from Eq.\,\ref{eq:f} ($\kappa/\beta c = 0.07\,$au using $B=10^{-5}$G). Thus, a continuous spatial point source injection close to the origin (at $\sim1.3\,R_\odot$ in our case) with a $p^{-2}$ profile in momentum should result in a steady-state solution with the same momentum power law of $p^{-2}$ at all radii until the cosmic rays escape from the spatial outer boundary. The blue dots in Fig.\,\ref{fig:appendix} represent the cosmic ray intensities as a function of kinetic energy from the simulation at $r\sim 1\,$au. The dashed line overplot a $p^{-2}e^{-p/p_\text{max}}/\beta$ profile for comparison and show that our results match well the expected result.

The second case (green dots in Fig.\,\ref{fig:appendix}) illustrates the effect of a spatially varying diffusion coefficient which also depends on momentum ($\kappa = \kappa(r,p)$, as is used in the simulations generally and using the magnetic field profile for the $\Omega = 3.5\Omega_\odot$ case). For a continuous spatial point source injection the particles now diffuse in a momentum-dependent way and the expected profile is $p^{-3}e^{-p/p_\text{max}}/\beta$. The green dashed line overplots a $p^{-3}e^{-p/p_\text{max}}/\beta$ profile for comparison and show that our results match well the expected result.

Finally, we include all three terms in the transport equation which is shown by the solid magenta line in Fig.\,\ref{fig:appendix}. In comparison to the diffusion only case, the cosmic ray fluxes are decreased at 1\,au by nearly 2 orders of magnitude due to spatial and momentum advection.

\section{Influence of the maximum momentum}
\label{subsec:pb}
\begin{figure}
        \includegraphics[width=0.5\textwidth]{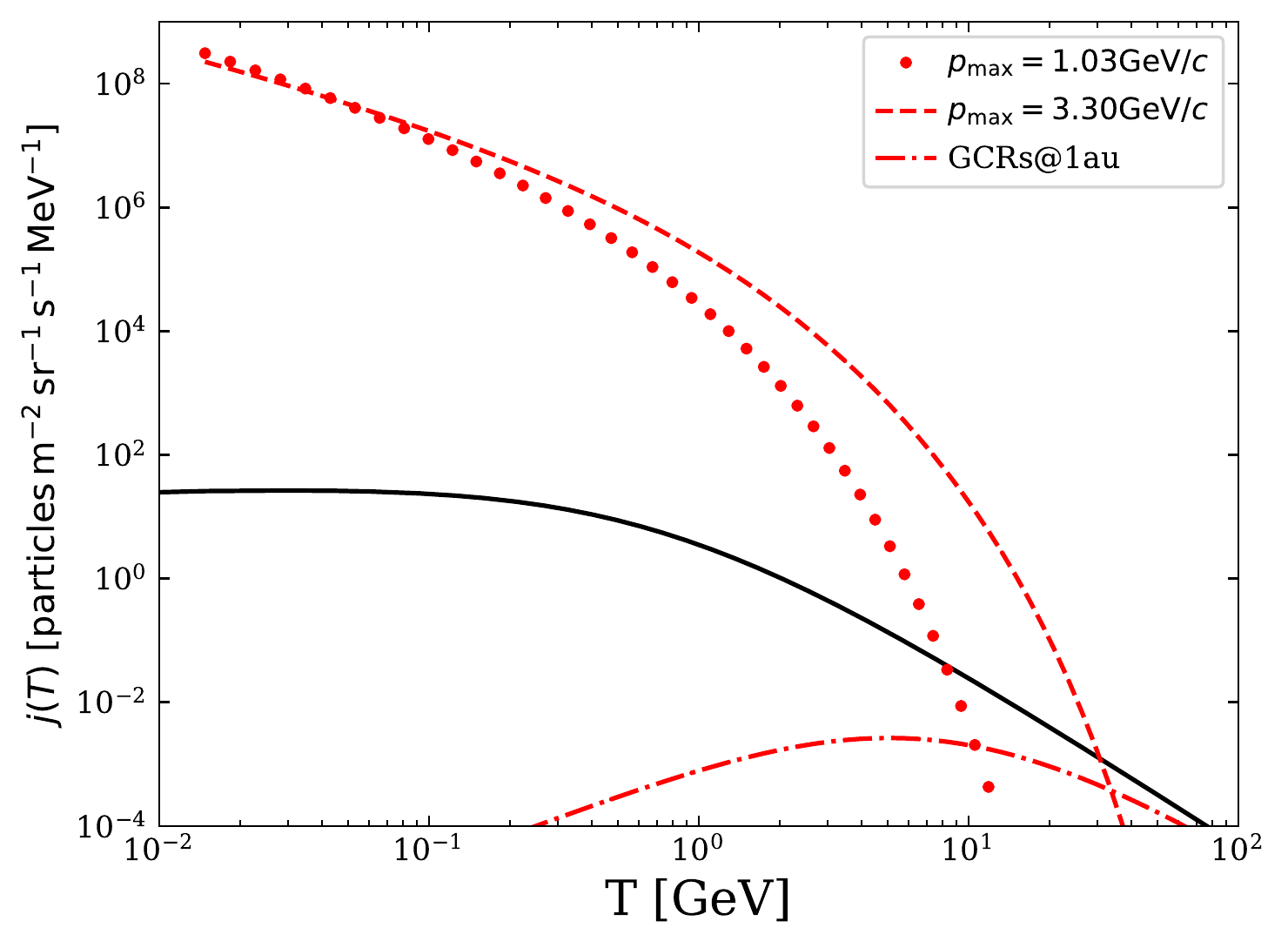}
	\centering
    \caption{The differential intensities for stellar (with two different spectral breaks) and Galactic cosmic rays at 1\,au are plotted for $\Omega=3.5\Omega_\odot$. The dotted lines represent the same values as in Fig.\,\ref{fig:junnormalised}.  }
    \label{fig:comparisons}
\end{figure}

Here, we investigate the sensitivity of our results on our choice of $p_\mathrm{max,\odot}$ which is used to normalise the scaling relation in Eq.\,\ref{eq:pb}. We increase $p_\mathrm{max,\odot}$ to $0.6\,\mathrm{GeV/c}$, increasing the maximum momentum to 3.3 $\mathrm{GeV}/c$ for $\Omega=3.5\Omega_\odot$. We compare the results of the simulation using this higher maximum momentum with the value adopted in the previous section. Fig.\,\ref{fig:comparisons} shows the differential intensities obtained from these simulations. The red dashed line in Fig.\,\ref{fig:comparisons} represents the results obtained using $p_\mathrm{max}=3.30\mathrm{GeV}/c$. The red dots are the same as the results shown in Fig.\,\ref{fig:junnormalised} using the lower value of $p_\mathrm{max}=1.03\mathrm{GeV}/c$. The red dash-dotted line represents the differential intensities for Galactic cosmic rays at 1\,au. The effect of changing the maximum momenta is quite significant. The higher spectral break means that stellar cosmic rays would dominate over Galactic cosmic ray fluxes up to $\sim$33\,GeV, in comparison to $\sim$10\,GeV for the lower spectral break.

This increase in the intensities occurs because of the timescales for the different physical processes (shown in Fig.\,\ref{fig:timescales_omega21} for $\Omega = 4\Omega_\odot$). By increasing the spectral break to $3.3\mathrm{GeV}/c$ there are sufficient numbers of $\gtrsim$GeV energy cosmic rays that can avoid  momentum losses in the innermost region of the stellar wind. 

\section*{Data Availability}
The output data underlying this article will be available via zenodo.org upon publication.

\newcommand\aj{AJ} 
\newcommand\actaa{AcA} 
\newcommand\araa{ARA\&A} 
\newcommand\apj{ApJ} 
\newcommand\apjl{ApJ} 
\newcommand\apjs{ApJS} 
\newcommand\aap{A\&A} 
\newcommand\aapr{A\&A~Rev.} 
\newcommand\aaps{A\&AS} 
\newcommand\mnras{MNRAS} 
\newcommand\pasa{PASA} 
\newcommand\pasp{PASP} 
\newcommand\pasj{PASJ} 
\newcommand\solphys{Sol.~Phys.} 
\newcommand\nat{Nature} 
\newcommand\bain{Bulletin of the Astronomical Institutes of the Netherlands}
\newcommand\memsai{Mem. Societa Astronomica Italiana}
\newcommand\apss{Ap\&SS} 
\newcommand\qjras{QJRAS} 
\newcommand\pof{Physics of Fluids}
\newcommand\grl{Geophysical Research Letters}
\newcommand\planss{Planetary and Space Science}
\newcommand\ssr{Space Science Reviews}
\newcommand\astrobiology{Astrobiology}
\newcommand\icarus{Icarus}
\newcommand\jgr{Journal of Geophysical Research}
\bibliographystyle{mn2e}
\bibliography{../../donnabib}

\label{lastpage}

\end{document}